\documentclass[prd,twocolumn,superscriptaddress,nofootinbib,showpacs]{revtex4}

\usepackage{graphicx,amsmath,amsfonts,amssymb,revsymb,dcolumn,epsfig,bm}

\newcommand{\bea}{\begin{eqnarray}}
\newcommand{\eea}{\end{eqnarray}}

\begin{document}

\title{G\"{o}del-type universes and chronology protection in Ho\v{r}ava-Lifshitz gravity}

\author{J.B. Fonseca-Neto}
\affiliation{Departamento de F\'{\i}sica, Universidade Federal da Para\'{\i}ba,\\
Caixa Postal 5008, 58051-970, Jo\~ao Pessoa -- PB, Brazil}

\author{A.Yu.\ Petrov}
\affiliation{Departamento de F\'{\i}sica, Universidade Federal da Para\'{\i}ba,\\
Caixa Postal 5008, 58051-970, Jo\~ao Pessoa -- PB, Brazil}

\author{M.J. Rebou\c{c}as}
\affiliation{Centro Brasileiro de Pesquisas F\'{\i}sicas,
Rua Dr.\ Xavier Sigaud 150,  \\
22290-180 Rio de Janeiro -- RJ, Brazil}

\date{\today}

\begin{abstract}
In the attempts toward a quantum gravity theory, general relativity faces a serious
difficulty since it is non-renormalizable theory.
Ho\v{r}ava-Lifshitz  gravity offers a framework to circumvent
this difficulty, by sacrificing  the local Lorentz invariance at ultra-high energy scales
in exchange of power-counting renormalizability. The Lorentz symmetry
is expected to be recovered at low and medium energy scales.
If gravitation is to be described by a Ho\v{r}ava-Lifshitz gravity theory there are a
number of issues that ought to be reexamined in its context, including the question as to whether
this gravity incorporates a chronology protection, or particularly if it allows G\"odel-type solutions
with violation of causality. We show that Ho\v{r}ava-Lifshitz gravity only allows hyperbolic G\"odel-type
space-times whose essential parameters $m$ and $\omega$ are in the chronology respecting 
intervals, excluding therefore any noncausal G\"odel-type space-times in the hyperbolic class.
There emerges from our results that the famous
noncausal G\"odel model is not allowed in Ho\v{r}ava-Lifshitz gravity.
The question as to whether this quantum gravity theory permits hyperbolic G\"odel-type solutions in the
chronology preserving interval of the essential parameters is also examined.
We show that Ho\v{r}ava-Lifshitz gravity not only excludes the noncausal G\"odel universe,
but also rules out any hyperbolic G\"odel-type  solutions  for physically
well-motivated perfect-fluid matter content.
\end{abstract}

\pacs{95.30.Sf, 98.80.Jk, 04.50.Kd, 95.36.+x}

\maketitle

\section{Introduction}

Even though general relativity is a highly successful classical field theory of gravity,
it faces a serious difficulty in the attempts toward a theory of quantum gravity since
one cannot quantize it by using the canonical quantization or path integral formalism
--- there emerges that it is a non-renormalizable theory.
Ho\v{r}ava-Lifshitz (HL) gravity~\cite{Hor} offers framework to circumvent
this difficulty, by sacrificing  the local Lorentz invariance at ultra-high energy scales
(typically trans-Planckian) in exchange of power-counting renormalizability.
The Lorentz symmetry is abandoned by invoking 
a different kind of scaling, called anisotropic or Lifshitz scaling~\cite{Lifshitz},
between space and time,  and it is expected that it is recovered at low and medium
(sub-Planckian) energy scales (long distance).

Since the publication of Ho\v{r}ava proposal in 2009~\cite{Hor}, a great deal of effort
has gone into the study of several features of Ho\v{r}ava-Lifshitz gravity. One can roughly
group the contributions to this issue into two broad families. In the first, one finds
articles devoted to the several aspects of  Ho\v{r}ava-Lifshitz gravity as a quantum field theory.
This class includes, among other matters, the attempts toward a consistent quantization of the
theory~\cite{MiaoLi,Rest,Kluson} and the calculation of counter-terms.%
\footnote{For some further references on several quantum aspects of Ho\v{r}ava-Lifshitz gravity
we refer the readers to Ref.~\cite{Add-ref-quant-aspects-HL}.}
In the second family, one has a number of interesting cosmological implications of
Ho\v{r}ava-Lifshitz gravity,  and the exam of some important solutions of Einstein's
equations in the framework of Ho\v{r}ava-Lifshitz gravity. This includes, for example,
Friedmann-Lema\^{\i}tre-Robertson-Walker models~\cite{Calc,Kiri,HorSol} and black-hole
solutions~\cite{Cai}, anisotropic scaling as a solution to the horizon problem
and as a way of having scale-invariant cosmological perturbations without
inflation~\cite{Mukohyama2009a}, dark matter as an integration constant~\cite{Mukohyama2009b},
and bounce solutions in the early universe~\cite{{Brandenberger2009}}.
For some further references on several cosmological implications of Ho\v{r}ava-Lifshitz gravity
see Ref.~\cite{Add-ref-cosmol-implications-HL} and references therein quoted on this issue.

Chronology and causality are central ingredients in the foundation of
the special relativity theory --- chronology is preserved and causality is respected.
The space-times of the general relativity  have locally the same causal
structure of the flat space-time of special relativity since   
a local chronology protection is inherited from the very fact that  
the space-times of general relativity are locally Minkowskian.
On nonlocal (global) scale, however, significant differences may arise since
Einstein's field equations do not provide nonlocal (topological) constraints on
the space-times.
Indeed, it has long been known that there are solutions to the general relativity
field equations that present causal anomalies in the form of closed time-like curves
(see, for example, Refs.~\cite{examples}).
The renowned model found by G\"odel~\cite{Godel49} is a well-known example of
a solution to Einstein's equations 
that makes it apparent that general relativity
permits solutions with closed timelike world lines, despite its local
Lorentzian character that leads to the local validity of the causality principle.
The G\"odel model is a solution of Einstein's equations with cosmological constant
$\Lambda$ for dust of density $\rho$, but it can also be interpreted as perfect-fluid
solution with equation of state $p=\rho\,$ without cosmological constant.
Owing to its unexpected features, G\"odel's model has a
recognizable importance and has motivated an appreciable 
number of investigations on rotating G\"odel-type models
as well as on causal anomalies not only in the context of general
relativity (see, e.g. Refs.~\cite{GT_in_GR})
but also in the framework of other theories of gravitation
(see, for example, Refs.~\cite{GT_Other_Th}).

The chronology protection conjecture introduced by Hawking~\cite{Hawking1992}
suggests that even though closed timelike curves are classically possible to be produced,
quantum effects are likely to prevent such time travel.  In this way, the laws of quantum
physics would prevent closed timelike curves from appearing.%
\footnote{For a good pedagogical overview with a fair list of references on the chronology
protection conjecture see  Visser~\cite{M_Visser_2003}.}

If gravitation can be described by  Ho\v{r}ava-Lifshitz gravity
theory there are a number of matters that ought to be reexamined in its framework,
including the question as to whether this quantum gravity theory permits
G\"odel-type solutions 
with violation of causality, somehow incorporating or not the chronology protection
conjecture for this family of spacetimes. Our chief aim in this paper  is to examine
this question by investigating the possibility of G\"odel-type universes along with
the question of breakdown of causality in Ho\v{r}ava-Lifshitz quantum gravity.%
\footnote{This extends the investigations on these issues carried out
in the framework of general relativity and other classical theories of
gravity (see, e.g., Refs.~\cite{Reb_Tiomno,RS-2009,SRO-2010}).}
We show that Ho\v{r}ava-Lifshitz gravity only allows hyperbolic G\"odel-type space-times
whose essential parameters $m^2$ and $\omega^2$ are in the chronology respecting
intervals, excluding therefore the noncausal G\"odel-type space-times in this class.
Thus, the famous noncausal G\"odel model is not
allowed in context of Ho\v{r}ava-Lifshitz gravity.
The question as to whether this quantum gravity theory permits hyperbolic G\"odel-type
solutions in the surviving chronology preserving interval of the essential parameters
is also examined.
We show that Ho\v{r}ava-Lifshitz gravity not only excludes the noncausal G\"odel model,
but also rules out any hyperbolic G\"odel-type solutions  for physically
well-motivated perfect-fluid matter content, which is the matter source for
the G\"odel universe in general relativity.

\section{G\"{o}del-type metrics}

It is well known that  G\"odel solution to the general relativity field equations
is a member of the following broad family of space-time-homogeneous (ST-homogeneous)
G\"odel-type geometries, whose form in cylindrical coordinates [$(r, \phi, z)$] is given
by~\cite{Reb_Tiomno}
\begin{equation}  \label{G-type_metric}
ds^2 = - [dt + H(r)d\phi]^2 + D^2(r)d\phi^2 + dr^2 + dz^2\,,
\end{equation}
where the functions $H(r)$ and $D(r)$ are such that
\begin{eqnarray}
\frac{H'}{D} & = & 2\omega\,,\label{eq:hcond_rot}\\
\frac{D''}{D} & = & m^{2},\label{eq:hcond_mass}
\end{eqnarray}
where the prime denotes derivative with respect to $r$, and the parameters $(m,\omega)$
are constants such that $\omega^{2}>0$ and  $-\infty\leq m^{2}\leq\infty$.

The ST-homogeneous G\"odel-type space-times can be grouped in the following classes:
\begin{enumerate}  
\item[\bf i.] Hyperbolic, in which $m^{2} = \mbox{const} > 0$ and
\begin{equation} \label{HD-hyperb}
H =\frac{4\,\omega}{m^{2}} \,\sinh^2 (\frac{mr}{2}), \;\;\;\;
               D=\frac{1}{m} \sinh\,(mr)\,;
\end{equation}

\item[\bf ii.] Trigonometric,  where $m^{2}= \mbox{const} \equiv - \mu^{2} < 0 $\  and
\begin{equation} \label{HD-circul}
 H = \frac{4\,\omega}{\mu^{2}} \,\sin^2 (\frac{\mu r}{2}),
                            \;\;\;\; D=\frac{1}{\mu} \sin\,(\mu r)\,;
\end{equation}
\item[\bf iii.] Linear, in which $m=0$ and
\begin{equation} \label{HD-linear}
 H = \omega r^{2},  \;\;\;\; D=r  \,.
\end{equation}
\end{enumerate}

We recall that in the above three families the constant $\omega$ is the vorticity of
matter source, and that all G\"odel-type metrics in the above classes are characterized
by the two essential parameters $\omega$ and $m$: identical pairs $(m^2, \omega^2)$ determine
isometric G\"odel-type space-times~\cite{Reb_Tiomno,TeiRebAman,RebAman}.
Moreover, G\"odel solution is just a particular case of the hyperbolic ($m^2 > 0$)  class
with $m^{2}= 2\,\omega^{2}$.

\section{Violation of causality and Ho\v{r}ava-Lifshitz gravity}

The causality features in Ho\v{r}ava-Lifshitz gravity can be looked upon
as having two interconnected physically significant ingredients, namely the gravity
theory, which involves the matter source, and the space-time geometry.
Regarding the latter, we begin by rewriting  the G\"odel-type line
element~(\ref{G-type_metric}) in the form
\begin{equation} \label{G-type_metric2}
ds^2=-dt^2 -2\,H(r)\, dt\,d\phi + dr^2 + G(r)\,d\phi^2 + dz^2 \,,
\end{equation}
where $G(r)= D^2 - H^2$.
In this form it is easy to show that existence of closed timelike curves,
which allows for violation of causality, depends upon the sign of  the
metric function $G(r)$.
Indeed, from Eq.~(\ref{G-type_metric2}) one has that the circles  defined
by $t, z, r = \text{const}$ become closed timelike curves whenever $G(r) < 0$.
Thus, the causality features of all ST-homogeneous G\"odel-type space-times
can be investigated by using essentially this inequality together with the basic 
variables and the field equations of Ho\v{r}ava-Lifshitz gravity.

Regarding the second ingredient in the  causality problem, we recall that
the basic quantities of  Ho\v{r}ava-Lifshitz gravity are the
lapse (real) function $N(t)$, the shift vector field $N^i(t,\vec{x})$ and the $3$-D
metric $g_{ij}(t,\vec{x})$ with which we write the spacetime metric in
the ADM form%
\footnote{Throughout this paper we use Greek letters to denote spacetime coordinate indices,
which are lowered and raised, respectively,  with $g_{\mu \nu }$ and $g^{\mu \nu }$, and vary from
$0$ to $3$,  whereas the spatial components $1 \cdots 3$ are denoted by Latin lower case letters
which are lowered and  raised with $g_{ij}$ and $g^{ij}$, respectively.}
\begin{equation} \label{ADM-form}
ds^{2}=-N^{2}dt^{2}+g{}_{ij}(dx^{i}+N^{i}dt)(dx^{j}+N^{j}dt)\,.
\end{equation}
{From }this equation together with Eq.~(\ref{G-type_metric2}) one has that the ADM variables
for the G\"odel-type metrics in cylindrical coordinates are given by  $N=D(r)\,/\sqrt{G(r)}$,
$N_{i}=(0,-H(r),0)$ and the spatial metric  
\begin{eqnarray}
g_{ij} & = & \left(\begin{array}{ccc}
1 & 0 & 0\\
0 & G(r) & 0\\
0 & 0 & 1
\end{array}\right)   \,.      \label{eq:adm metr}
\end{eqnarray}

For the hyperbolic family of G\"odel-type metrics, from Eq.~(\ref{HD-hyperb}) one finds that
\begin{equation}
G(r) = \frac{4}{m^2} \, \sinh^2 (\frac{mr}{2}) \left[ ( 1- \frac{4\omega^2}{m^2})\,
\sinh^2 (\frac{mr}{2})+1 \right]\,,
\end{equation}
and therefore for $0 < m^2 < 4\omega^2$ there is a critical radius $r_c$ defined
by $G(r)=0$, namely
\begin{equation} \label{r-critical}
\sinh^2 \frac{mr_c}{2}= \left[ \frac{4\omega^2}{m^2} - 1 \right]^{-1}\,,
\end{equation}
such that $G(r)>0$ for $r<r_{c}$ and $G(r)<0$ for $r>r_{c}$. Hence, the circles
$t,r,z=\text{const}$  in the circular band with  $r>r_{c}$ are closed
timelike curves.%
\footnote{We note that the only G\"odel-type metric without such noncausal circles
comes about when $m^2 = 4 \omega^2$ (see Ref.~\cite{Reb_Tiomno}).
In this case, the critical radius $r_c \rightarrow \infty$,
and hence the violation of causality of G\"odel type is avoided.}
However, on the one hand the Riemannian (positive definite) character of the spatial metric $g_{ij}$
implies that $g=\det(g_{ij})=\sqrt{G(r)}>0$, on the other hand  the lapse is also an
imaginary function in the noncausal region $G(r)<0$ defined by $t, z, r = \text{const}$ and $r>r_c$.
Thus, for the hyperbolic class,
the noncausal space-times are excluded in Ho\v{r}ava-Lifshitz gravity.
Therefore, from this result one has that the famous
G\"odel model, for which $m^2 = 2\omega^2$, is not permitted in Ho\v{r}ava-Lifshitz gravity.
%
\footnote{We note that in the parameter interval $m^2 > 4\omega^2$ one has $G(r)>0$. Thus,
$m^2 > 4\omega^2$ defines the causal parameter interval in  G\"odel-type class of spacetimes,
which is permitted in Ho\v{r}ava-Lifshitz context.}

A similar analysis holds for the remaining classes of G\"odel-type space-times.
Indeed, for the trigonometric class whose metric functions are given by Eq.~(\ref{HD-circul})
one finds that%
\begin{equation} 
G\left(r\right)=\frac{4\,}{\mu^{4}}\sin^{2}(\frac{\mu\, r}{2})\,[\,\mu{}^{2}
-(4\,\omega^{2}+\mu{}^{2})\, \sin^{2}(\frac{\mu\, r}{2})\,] \,,\label{eq:G trig}
\end{equation}
and therefore  $G(r)$  has an infinite sequence of zeros. Thus, there is an infinite
sequence of alternating causal [$\,G(r) > 0\,$] and noncausal [$\,G(r) < 0\,$] regions 
in the section $t, z, r = \text{const}$, without and with noncausal circles, depending on
the value of $r=\text{const}$ (see Appendix for detailed calculations).
Thus, for example, if $G(r) < 0$ for a certain range of $r$ ($ r_1 < r < r_2$, say)
noncausal G\"odel's circles exist, whereas  for $r$ in the next circular band $r_2 < r < r_3$ (say)
for which $G(r) > 0$ no such noncausal circles exist. Nevertheless, since in
context of Ho\v{r}ava-Lifshitz gravity the spatial metric $g_{ij}$ is positive definite,
the regions of the underlying G\"odel-type manifolds in which the chronology is violated,
i.e. $t, z, r = \text{const}$ with $G(r) < 0$,  are excluded for the trigonometric family
of spacetimes. In these regions the lapse function again becomes an imaginary function.

Finally, for the linear family,  from Eq.~(\ref{HD-linear}) one
easily finds
\begin{equation} 
G(r)=r^{2}-r^{4}\,\omega^{2}=-r^{2}\,\left(r\,\omega-1\right)\,
\left(r\,\omega+1\right) \,.\label{eq:G linear} 
\end{equation}
Thus, there is a critical radius [$\,G(r)=0\,$] given  by $r_c = 1/\omega$,
such that for any radius $r > r_c$ the inequality $G(r)<0$ holds, and then the
circles defined by $t, z, r = \text{const}$ are closed timelike curves. 
Here again the positive definite character of spatial metric and the fact the lapse is a
real function cannot be imposed in the regions of G\"odel-type manifolds that violate the
chronology [$G(r)<0$]. Thus, the noncausal region of the linear family is excluded in
Ho\v{r}ava-Lifshitz gravity.

To summarize the above results, we have shown that Ho\v{r}ava-Lifshitz  gravity can only
be consistently formulated in the chronology preserving regions of G\"odel-type manifolds,
excluding therefore the noncausal regions of underlying G\"odel-type manifolds for  
all classes of ST-homogeneous G\"odel-type spacetimes.
This excludes any G\"odel-type space-times with violation of causality. Particularly
for the hyperbolic class it rules out the well-known G\"odel metric for which
$m^2 = 2 \omega^2$. As a matter of fact, since we have not used so far the
Ho\v{r}ava-Lifshitz field equations, this result holds for any theory whose
formulation relies on the suitable behavior of the ADM variables of the
G\"odel-type  space-times.

The fact that Ho\v{r}ava-Lifshitz gravity does not permit hyperbolic G\"odel-type metrics
whose essential parameters $m$ and $\omega$ define noncausal G\"odel-type space-time
geometries does not signify that space-times such as wormholes,
which are seem generically to lead to the creation of time machines, cannot be
found in Ho\v{r}ava-Lifshitz gravity~\cite{HL-Wormhole2010}.
Moreover, although the presence of a single closed timelike curve as,
for example,  the above G\"odel's circles ($t, z, r = \text{const}>r_{c}$),
is an unequivocal manifestation of violation of the chronology protection conjecture,
a space-time may admit noncausal closed curves other than these G\"odel's circles.
This means that the exclusion of all noncausal hyperbolic G\"odel space-times can 
only be seen as a tiny suggestion that the chronology is protected in Ho\v{r}ava-Lifshitz gravity
in the sense that it is protected for this type of causal anomaly of G\"odel-type
space-times.

Given that the noncausal interval of the essential parameters of hyperbolic G\"odel-type
spacetimes are excluded in Ho\v{r}ava-Lifshitz context, a question arises as to whether 
this theory permits solutions of its field equations in the  region  where the
chronology is respected [$\,G(r)>0\,$] for physically well-motivated matter content.
In the next section we shall examine this question for a perfect-fluid
matter source.

\section{Ho\v{r}ava-Lifshitz gravity}

\subsection{Field equations}

Here  we  briefly introduce the Ho\v{r}ava-Lifshitz gravity and present its
field equations in the form that will be used in the next section.
We begin by recalling that the dynamical variables of this theory are
the lapse function $N(t)$, the shift vector field $N^i(t,\vec{x})$ and the spatial
metric $g_{ij}(t,\vec{x})$ with which we rewrite an arbitrary spacetime line
element
\begin{equation}
ds^2=g_{00}\,dt^2 + 2\,g_{0i}\, dx^idt + g_{ij}\, dx^idx^j
\end{equation}
in the ADM form given by Eq.~(\ref{ADM-form}).
Thus, the ADM variables can be expressed in terms of the metric components $g_{\mu\nu}$
as $N_i=g_{0i}$ and $N=(g_{ij}N^iN^j-g_{00})^{1/2}$.

The Lagrangian for the Ho\v{r}ava-Lifshitz gravity we consider in this paper
is given by~\cite{HorSol,Kiri}
\begin{eqnarray} \label{lagra}
L=\sqrt{g}N\Big[\frac{2}{\kappa^2}\Big(K_{ij}K^{ij}-\lambda K^2\Big)-\frac{\kappa^2}{2w^4}C_{ij}C^{ij}+
&&\nonumber\\
+\frac{\kappa^2\mu}{2w^2}\frac{\epsilon^{ijk}}{\sqrt{g}}R_{il}\nabla_jR^l_k
-\frac{\kappa^2\mu^2}{8}R_{ij}R^{ij} && \nonumber \\
+\frac{\kappa^2\mu^2}{8(1-3\lambda)}\Big(\frac{1-4\lambda}{4}R^2+\Lambda R-3\Lambda^2\Big)
+{\cal L}_m\Big], &&
\end{eqnarray}
where
\bea \label{extr-curv}
K_{ij}=\frac{1}{2N}(\dot{g}_{ij}-\nabla_i N_j-\nabla_j N_i)
\eea
is the extrinsic curvature, overdot stands for derivative with respect to $t$,
$K=g^{ij}K_{ij}$ is its trace, $R_{ij}$ is the Ricci tensor for the metric $g_{ij}$,
\bea  \label{cotton-ten}
C^{ij}=\frac{\epsilon^{ikl}}{\sqrt{g}}\nabla_k(R^j_l-\frac{1}{4}R\delta^j_l)
\eea
is the Cotton tensor, and  ${\cal L}_m$ is the matter Lagrangian, which
depends on the matter fields and on the ADM variables. In Eq.(\ref{lagra}),
 $\Lambda$ is a cosmological constant, $\kappa^2$ is a gravitational constant,
and $\lambda$, $w$, $\mu$ are coupling parameters of the theory.
It should be noticed that if one keeps the spatial derivative only up to the
second order, for $\lambda=1$ the general relativity is recovered.
We also note that the Lagrangian~(\ref{lagra}) involves terms with different
values of the critical exponent $z$. To recover general relativity the $z=1$ terms
are necessary, whereas $z=3$ terms are needed for renormalizability.

The equations of motion can now be obtained through the variation of the action defined by the
Lagrangian~(\ref{lagra}) with respect to the ADM dynamical variables. Indeed,

\textbf{(i)}
variation with respect to  $N$  and  $g_{00}$ are related and the result is given by~\cite{Kiri}
\bea
\label{eqG00}
\frac{\delta S}{\delta g_{00}} &=&
\Big(\frac{\delta S_{g}}{\delta N}+\frac{\delta S_{m}}{\delta N}\Big)
\frac{\delta N}{\delta g_{00}}   = G^{00}-T^{00}=0,
\eea
where $\frac{\delta N}{\delta g_{00}}=-\frac{1}{2N}$, $G^{\mu\nu}$ is the generalized Einstein tensor with
\bea
\label{tens1}
G^{00}&=&\frac{1}{2N} \Big[\,-\alpha(K_{ij}K^{ij}-\lambda K^2)+\beta C_{ij}C^{ij}  \\  \nonumber
&+& \sigma +\gamma\frac{\epsilon^{ijk}}{\sqrt{g}}R_{il}\nabla_jR^l_k
+\zeta R_{ij}R^{ij}+\eta R^2+\xi R \,\Big]
\eea
and $T^{\mu\nu}$ is the energy-momentum tensor of matter;

\textbf{(ii)}
variation with respect to $N_l=g_{0l}$ furnishes~\cite{Kiri}
\bea
\label{eqG0i}
\frac{\delta S}{\delta N_l}=G^{0l}-T^{0l}=2\alpha\nabla_k(K^{kl}-\lambda Kg^{kl})-T^{0l}=0;
\eea
\textbf{(iii)}
variation with respect to $g_{ij}$ provides~\cite{Kiri}
\bea
\label{eqGij}
G_{ij}=T_{ij}\,,
\eea
where
\bea
\label{ein}
G_{ij}=G^{(1)}_{ij}+G^{(2)}_{ij}+G^{(3)}_{ij}+G^{(4)}_{ij}+G^{(5)}_{ij}+G^{(6)}_{ij}
\eea
and
\begin{widetext}
\bea
\label{eincomp}
G^{(1)}_{ij}&=&2\alpha NK_{ik}K_j^k-\frac{\alpha N}{2}K_{kl}K^{kl}g_{ij}+ 
\alpha(K_{ik}N_j)^{;k}+\alpha(K_{jk}N_i)^{;k}-\alpha(K_{ij}N_k)^{;k}+(i\leftrightarrow j)\,,
\nonumber\\
G^{(2)}_{ij}&=&-2\alpha \lambda NKK_{ij}+\frac{\alpha\lambda
N}{2}K^2g_{ij}-\frac{\alpha\lambda}{\sqrt{g}}g_{ik}g_{jl}\frac{\partial}{\partial t}(\sqrt{g}Kg^{kl})
 \nonumber \\&-&
\alpha\lambda(Kg_{ik}N_j)^{;k}-\alpha\lambda(Kg_{jk}N_i)^{;k}+\alpha\lambda(Kg_{ij}N_k)^{;k}+(i\leftrightarrow j)\,,
\nonumber\\
G^{(3)}_{ij}&=&N\xi R_{ij}-\frac{N}{2}(\xi R+\sigma)g_{ij}-\xi N_{;ij}+\xi \Box N g_{ij}+(i\leftrightarrow j)\,,
\nonumber\\
G^{(4)}_{ij}&=&2N\eta RR_{ij}-\frac{N}{2}\eta R^2g_{ij}+2\eta \Box(N R)g_{ij}-2\eta(NR)_{;ij}+(i\leftrightarrow j)\,,
\nonumber\\
G^{(5)}_{ij}&=&\Box (N (\zeta R_{ij}+\frac{\gamma}{2}C_{ij})) -(N(\zeta R_{ki}
+\frac{\gamma}{2}C_{ki}))_{;j}^{;\phantom{j}k}
+ (N (\zeta R^{kl}+\frac{\gamma}{2}C^{kl}))_{;lk}\,g_{ij}+ (i\leftrightarrow j)\,,
 \nonumber\\&&
\nonumber\\
G^{(6)}_{ij}&=&\frac{1}{2}\frac{\epsilon^{mkl}}{\sqrt{g}}\Big[(Q_{mi})_{;kjl}+(Q_m^{\phantom{k}n})_{;kin}g_{jl}-
(Q_{mi})_{;kn}^{;\phantom{kk}n}g_{jl}-(Q_{mi})_{;k}R_{jl}-
(Q_{mi}R_k^n)_{;n}g_{jl}  \nonumber\\&+&
(Q_{\phantom{n}m}^nR_{ki})_{;n}g_{jl}+\frac{1}{2}(R^n_{\phantom{n}pkl}Q^{\phantom{p}p}_m)_{;n}g_{ij}+Q_{mi}R_{jl;k}
\Big]+2N\zeta R_{ik}R_j^k  \nonumber \\&-&
\frac{N}{2}(\beta C_{kl}C^{kl}+\gamma R_{kl}C^{kl}+\zeta R_{kl}R^{kl})g_{ij}-\frac{1}{2}Q_{kl}C^{kl}g_{ij}
+(i\leftrightarrow j)\;.
\eea
\end{widetext}

In the above field equations we have defined
\bea
\label{def}
&&\alpha=\frac{2}{\kappa^2}\,,\quad\,\beta=-\frac{\kappa^2}{2w^4}\,,
\quad\,\gamma=\frac{\kappa^2\mu}{2w^2}\,,\quad\,\zeta=-\frac{\kappa^2\mu^2}{8}; \nonumber\\
\nonumber \\
&&\eta=\frac{\kappa^2\mu^2(1-4\lambda)}{32(1-3\lambda)},
\quad \xi=\frac{\kappa^2\mu^2\Lambda}{8(1-3\lambda)}, \qquad \tau = \frac{1}{1-3\lambda}\,, \nonumber \\
&&
\sigma=-\frac{3\kappa^2\mu^2\Lambda^2}{8(1-3\lambda)}\,, \quad Q_{ij}\equiv  N(\gamma R_{ij}+2\beta C_{ij})\,. 
\eea

\subsection{Perfect fluid as source}

Given that Ho\v{r}ava-Lifshitz gravity rules out  the chronology violating
of hyperbolic G\"odel-type spacetimes, the question as to whether
this theory admits  G\"odel-type solutions in the chronology preserving
interval of the essential parameters, $\,m^2 > 4\omega^2$, naturally arises here.
In this section we shall examine this question by considering
the hyperbolic class ($m^2 >0 $) of G\"odel-type [ Eq.~(\ref{HD-hyperb}) ]
in  the framework of Ho\v{r}ava-Lifshitz gravity.
This is the most important class of G\"odel-type spacetime geometries
as it contains the two most relevant G\"odel-type solutions of Einstein's
equations, namely G\"odel solution~\cite{Godel49},  
in which $m^2= 2\omega^2$, and the only causal G\"odel-type solution found
in Ref.~\cite{Reb_Tiomno}, in which $m^2 = 4\omega^2$.

To simplify the calculation of the geometrical quantities of the hyperbolic class
that are required for the Ho\v{r}ava-Lifshitz field equations,
we begin by introducing new (Cartesian) coordinates $t', x, y, z'$ defined through
the following coordinate transformation
\begin{eqnarray}
&& \tan [\,\phi/2 + (m^2/4\omega)\, (t'-t)\,]  =  e^{-mr} \tan (\phi/2),  \\
&& e^{mx} = \cosh (mr) +  \sinh (mr)\,\cos \phi \,,  \\
&& m\,y\,e^{mx} =  \sinh (mr)\, \sin \phi \,,  \\
&& z' =  z \,,
\end{eqnarray}
and rewrite the line element of the hyperbolic family given by Eq.~(\ref{HD-hyperb})
in the form
\begin{equation} \label{G-type_Cartesian}
ds^2 = - [dt' + (2\omega/m)\,e^{mx} dy]^2 + e^{2mx} dy^2 + dx^2 + dz'^2\,,
\end{equation}
where $-\infty < t', x, y, z' < + \infty$.  In this coordinates the field equations
for this class of G\"odel-type metrics become much simpler. We emphasize, however,
that the following results hold for to the whole hyperbolic ($m^2>0$)
class of G\"{o}del-type metrics.

{}From Eq.(\ref{G-type_Cartesian}) one has that the ADM variables are given
by  $N_{i}=(0,-(2\omega/m)\,e^{mx},0)$, $N=1/{v}$, and $g_{ij}=\text{diag}\,(1,G(x),1)$, where
\begin{equation} \label{vdef}
G(x)=v^{2}e^{2mx} \quad \text{with} \quad v=\sqrt{1-\left(\frac{2\omega}{m}\right)^{2}}\;.
\end{equation}

A straightforward calculation shows that the spatial metric $g_{ij}$ gives rise to
the following non-zero components of the Christoffel symbols $\Gamma_{\:12}^{2}=m$
and $\Gamma_{\:22}^{1}=-mv^{2}e^{2\, m\, x}$, from which one has the nonvanishing
component of the Riemannian curvature $R_{1212}=-m^{2}v^{2}\, e^{2\, m\, x}$.
Thus the corresponding components of the Ricci tensor  and the curvature scalar
are given, respectively, by $R_{11}=-m^{2}$ and $R_{22}=-m^{2}v^{2}\, e^{2\, m\, x}$,
$R=-2m^{2}$.
Now, by using this Ricci tensor and scalar,  it is easy to show that related
Cotton tensor (see Eq.~\ref{cotton-ten}) is identically null, $C^{ij}=0$.
On the other hand, the only nonvanishing component of the extrinsic curvature~(\ref{extr-curv})
is $ K_{12}=-v\,\omega e^{mx}$, which gives $K=g^{ij}K_{ij}=0$. This completes the
calculations of the geometrical quantities of hyperbolic  G\"odel-type geometries,
which are needed to have Ho\v{r}ava-Lifshitz field equations.

The other important ingredient of Ho\v{r}ava-Lifshitz field equations is the matter source.
Similarly to the G\"odel solution in the general relativity framework~\cite{Godel49}, we
consider in this work a perfect fluid of density $\rho$ and pressure $p$.  T
Thus, we have
\begin{equation}  \label{eq:em_ten_p0}
T^{\mu\,\nu}=(\rho+p)\, u^{\mu}\, u^{\nu}+p\, g^{\mu\,\nu}\,.
\end{equation}

Without loss of generality from now on we choose units such that $\kappa^2=1$.
Taking into account~(\ref{eq:em_ten_p0}), the field equations (\ref{eqG00}), (\ref{eqG0i})
and (\ref{eqGij}) reduce to the following set of  algebraic  equations: 
\begin{eqnarray}
-4\, m^{4}\,\tau\, v\,\zeta+2\,\Lambda\, m^{2}\,\tau\, v\,\zeta+3\,\Lambda^{2}\,\tau\, v\,\zeta\nonumber \\
+2\, m^{4}\, v\,\zeta+2\, p\, v^{2}-4\,\omega^{2}\, v-2\,\rho-2\, p & = & 0,\label{eq00}\\
2\,\omega\,(p\, v-4\, m^{2}) & = & 0,\label{eq02}\\
-4\, m^{4}\,\tau\,\zeta-3\,\Lambda^{2}\,\tau\,\zeta+2\, m^{4}\,\zeta-p\, v+4\,\omega^{2} & = & 0,\label{eq11}\\
-4\, m^{4}\,\tau\, v\,\zeta-3\,\Lambda^{2}\,\tau\, v\,\zeta+2\, m^{4}\, v\,\zeta\nonumber \\
+\rho\, v^{2}-12\,\omega^{2}\, v-\rho-p & = & 0,\label{eq22}\\
4\, m^{4}\,\tau\,\zeta-2\,\Lambda\, m^{2}\,\tau\,\zeta-3\,\Lambda^{2}\,\tau\,\zeta\nonumber \\
-2\, m^{4}\,\zeta-p\, v-4\,\omega^{2} & = & 0,\label{eq33}
\end{eqnarray}
written in terms of independent parameters $\rho,\Lambda,\tau,\zeta,\omega$,
and  $m^2$.

We recall that to find a G\"odel-type solution to the above algebraic
equations~(\ref{eq00})--(\ref{eq33}) amounts to determining a pair $(m^2 ,\omega^2)$
in the chronology preserving interval $m^2 > 4\omega^2$. In what follows we shall show
that such a pair does not exist, ruling out therefore any G\"odel-type solution for
a perfect fluid matter source in the hyperbolic class.  To this end, we
we first solve Eq.~(\ref{eq02}) for $p$ to have
\begin{equation}
p=\frac{4\, m^{2}}{v},
\label{psol}
\end{equation}
and then substitute the result back into the remaining field equations~(\ref{eq00}),
(\ref{eq11})--(\ref{eq33}), to obtain that
\begin{eqnarray}
4\, m^{4}\,\tau\, v\,\zeta-2\,\Lambda\, m^{2}\,\tau\, v\,\zeta-3\,\Lambda^{2}\,\tau\, v\,\zeta\nonumber \\
-2\, m^{4}\, v\,\zeta+4\,\omega^{2}\, v-8\, m^{2}\, v+\frac{8\, m^{2}}{v}+2\,\rho & = & 0,\label{eq00p}\\
4\, m^{4}\,\tau\,\zeta+3\,\Lambda^{2}\,\tau\,\zeta
-2\, m^{4}\,\zeta-4\,\omega^{2}+4\, m^{2} & = & 0,\label{eq11p}\\
4\, m^{4}\,\tau\, v^{2}\,\zeta+3\,\Lambda^{2}\,\tau\, v^{2}\,\zeta-2\, m^{4}\, v^{2}\,\zeta\nonumber \\
-\rho\, v^{3}+12\,\omega^{2}\, v^{2}+\rho\, v+4\, m^{2} & = & 0,\label{eq22p}\\
4\, m^{4}\,\tau\,\zeta-2\,\Lambda\, m^{2}\,\tau\,\zeta-3\,\Lambda^{2}\,\tau\,\zeta\nonumber \\
-2\, m^{4}\,\zeta-4\,\omega^{2}-4\, m^{2} & = & 0.\label{eq33p}
\end{eqnarray}

Now, we solve~(\ref{eq11p}) and~(\ref{eq33p}) for $\tau$ and $\zeta$ and find
\begin{eqnarray}
\tau & = & \frac{2\, m^{6}}{\left(\Lambda\, m^{2}+3\,\Lambda^{2}\right)\,\omega^{2}+4\, m^{6}
-\Lambda\, m^{4}}   \,,\label{eqtau}\\
\zeta & = & -\frac{\left(2\,\Lambda\, m^{2}+6\,\Lambda^{2}\right)\,\omega^{2}+8\, m^{6}
-2\,\Lambda\, m^{4}}{\Lambda\, m^{6}+3\,\Lambda^{2}\, m^{4}} \,,\label{eqzeta}
\end{eqnarray}
which can be substituted into~(\ref{eq00p}) and~(\ref{eq22p}), in
order to write the remaining two equations in the form
\begin{eqnarray}
\rho\, v-\frac{16\,\omega^{4}}{m^{2}}+12\,\omega^{2}+2\, m^{2} & = & 0\,,
\label{eq00pzt}\\
\rho\, v-16\,\omega^{2}+8\, m^{2} & = & 0\,.
\label{eq22pzt}
\end{eqnarray}
{}From Eq.~(\ref{eq22pzt}) we have
\begin{equation}
\rho=\frac{8\,\left(2\,\omega^{2}-m^{2}\right)}{v} \,,\label{rhosol}
\end{equation}
This equation together with Eq.~(\ref{eq00pzt}) gives
\begin{equation}
\left(4\,\omega^2-m^2\right)\,\left(2\,\omega^{2}-3\,m^{2}\right)=0\label{eqmw}
\end{equation}
whose solutions are
\begin{eqnarray} \label{msol}
m^2 = \frac{2}{3} \omega^2 \quad \text{and} \quad m^2 = \frac{1}{4}\omega^2\,,
\end{eqnarray}
which are both outside the chronology preserving interval $m^2 > 4\omega^2$,
making apparent that there is no perfect-fluid G\"odel-type solution to the
Ho\v{r}ava-Lifshitz field equations in the chronology preserving
region.%
\footnote{Note, in addition, that for the solutions~(\ref{msol}) one has,
respectively,  $v=0$ and $v=\sqrt{5}\,i$, which gives that $p$ and $\rho$ are either
undefined or imaginary quantities. This reinforce the fact that these solutions
are not permitted in the framework of Ho\v{r}ava-Lifshitz gravity.}
Furthermore, this result holds regardless of the equation of state $ p / \rho $.

To close this section, some words of clarification regarding the results of
Ref.~\cite{Furtando-etal2011} are in order. First, we note that rather than
dealing with the whole hyperbolic family of G\"odel-type space-times in this
reference only  the particular case of  G\"odel metric has been considered.
Second, we emphasize that their whole calculations were made
without noticing, for example, that lapse $N$ is not well-defined for particular
case of G\"odel metric. Thus, Ho\v{r}ava-Lifshitz gravity was improperly used in
the  chronology violating region  to define an energy-momentum tensor associated
to the G\"odel metric.
Indeed, since  $m^2 = 2\omega^2$ for G\"odel metric, Eq.~(30)
makes clear that, e.g.,  the lapse $N=1/v$,  with $v$ given by Eq.~(\ref{vdef}),
is an imaginary function. An important outcome of the above results is that the
famous G\"odel space-time cannot be a solution of Ho\v{r}ava-Lifshitz gravity
no matter how exotic is the source one takes~\cite{Furtando-etal2011}, since some
dynamical variables can only be consistently defined in the chronology preserving
for the range $m^2 > 4 \omega^2$ of G\"odel-type classes of manifolds.

\section{Concluding Remarks}

Despite its great success as a classical theory of gravity, general relativity faces
a crucial difficulty in the attempts toward a quantum theory of gravity in that
it is non-renormalizable. Hence, general relativity is viewed as an effective
theory that breaks down at some energy scale, beyond which it is unsuitable to describe
the gravitational interaction. Ho\v{r}ava-Lifshitz gravity evades this difficulty by
invoking an anisotropic scaling between space and time, which amounts to sacrificing
the local Lorentz invariance at ultra-high energy scales in exchange of power-counting
renormalizability. The Lorentz symmetry is expected to be recovered at low and medium energy
scales (long distance).

Chronology and causality are central ingredients in the foundation of the special
relativity theory. These properties are naturally inherited locally by general relativity
theory, whose  space-times are locally Minkowskian. The nonlocal question, however, is left
open, and violation of causality can come about. Indeed, it has long been known that
there are solutions of Einstein's equations that exhibit closed time-like
curves. The  G\"odel model is the best known example of a cosmological solution
of Einstein's  equations in which causality is violated at a nonlocal scale. 
In 1992 Stephen Hawking suggested that even though closed timelike curves %
can arise in the framework of classical theories of gravitation, quantum effects are
likely to prevent chronological pathologies. In this way, the laws of quantum
physics would prevent closed timelike curves from appearing. 

In this paper we proceeded further with the  investigations on the potentialities,
difficulties, and limitations of Ho\v{r}ava-Lifshitz gravity by investigating
the possibility of G\"odel-type solutions to its field equations along
with the question of breakdown of causality in Ho\v{r}ava-Lifshitz quantum gravity.
We have shown that Ho\v{r}ava-Lifshitz gravity only allows the chronology respecting
interval of the essential parameters of hyperbolic G\"odel-type spacetimes, excluding therefore
the noncausal hyperbolic G\"odel-type space-times.
Thus, there emerges from our results that the
well-known  noncausal G\"odel model is not permitted in context of Ho\v{r}ava-Lifshitz
gravity regardless of matter source, since some  ADM dynamical variables
can only be consistently defined in the chronology preserving parameter interval $m^2 > 4 \omega^2$
of hyperbolic G\"odel-type  space-time family.
As a consequence, G\"odel metric ($m^2 = 2 \omega^2 $) cannot be suitably used to define an
energy-momentum tensor through Ho\v{r}ava-Lifshitz field equations.
This illustrates concretely that the existence of a preferred foliation of
space-time brings on a distinctive causal structure in the context of Ho\v{r}ava-Lifshitz gravity.
Such a special causal structure puts the violation of causality of the general relativity theory 
into a new perspective.
It should be noted that since we have not used the Ho\v{r}ava-Lifshitz field equations to
derive this result, it holds for any theory whose formulation relies on the suitable 
behavior of the ADM variables of the G\"odel-type  space-times.
However, the fact that these gravity theories do not permit noncausal G\"odel-type
whose essential parameters $m$ and $\omega$ define noncausal hyperbolic G\"odel-type 
space-time geometries does not signify that space-times such a wormhole,
which are seem generically to lead to the creation of time machines, cannot be
found in Ho\v{r}ava-Lifshitz gravity~\cite{HL-Wormhole2010}.
This means that the exclusion of all noncausal hyperbolic G\"odel space-times can 
only be seen as a tiny suggestion that the chronology is protected in Ho\v{r}ava-Lifshitz 
gravity.
The question as to whether Ho\v{r}ava-Lifshitz gravity theory allows hyperbolic G\"odel-type 
solutions in the chronology preserving region of the essential parameters was also examined. 
We have shown that Ho\v{r}ava-Lifshitz gravity not only excludes the noncausal G\"odel model, 
but also rules out any G\"odel-type solutions of the hyperbolic class for physically well-motivated 
perfect-fluid matter content, which can be taken as the matter source for the G\"odel universe 
in the general relativity theory.

\begin{acknowledgments}
M.J. Rebou\c{c}as acknowledges the support of FAPERJ under a CNE E-26/101.556/2010 grant.
This work was also supported by Conselho Nacional de Desenvolvimento
Cient\'{\i}fico e Tecnol\'{o}gico (CNPq) - Brasil, under grant No. 475262/2010-7.
M.J. Rebou\c{c}as thanks  CNPq for the grants under which this work
was carried out. We are grateful to A.F.F. Teixeira for reading the manuscript
and indicating some omissions and typos. A.Yu.\ Petrov thanks J.R. Nascimento and
A.F. Santos for interesting discussions.
\end{acknowledgments}

\appendix*
\section{}
In this Appendix, we present the detailed calculations of the causality problem for
the trigonometric class of G\"odel-type space-times, which exhibits an infinite
sequence of alternating causal [$\,G(r) > 0\,$] and noncausal [$\,G(r) < 0\,$] regions
in the section $t, z, r = \text{const}$, without and with noncausal circles, depending on
the value of $r=\text{const}$. To this end, all we have to do is to determine the
behavior of the function $\,G(r)\,$ given by equation (\ref{eq:G trig}).
We begin by noting that this function has an infinite sequence of zeros
$G(r_{n})=0 \; (n=0,1,2,\ldots)$  determined by the equations
\begin{equation}
\sin\,(\frac{\mu\, r_n}{2}) = 0  \label{cond1Gzero}
\end{equation}
and
\begin{equation}
\sin(\frac{\mu\, r_n}{2})=\pm\frac{\mu(-1)^n}{\sqrt{4\,\omega^{2}+\mu{}^{2}}}\,,
\label{cond2Gzero}
\end{equation}
%
\begin{figure}[tb!]
\includegraphics[width=6.7cm,height=4.2cm,angle=0]{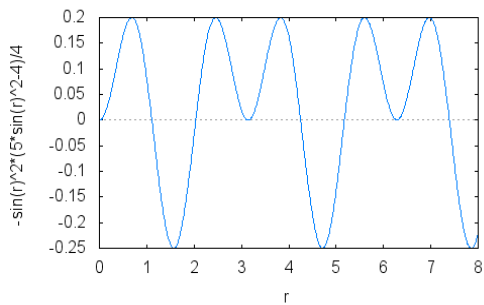}  
\caption{\label{fig1} This figure illustrates a typical behavior of the function
$G(r)$ for the trigonometric class of G\"odel-type spacetimes.}
\end{figure}
%
whose roots are given by
\begin{equation}
r_1^{(n)}=\frac{2\,\pi\, n}{\mu},\: n=0,1,2,\ldots
\label{eq:G trig zero r1n}
\end{equation}
for Eq.~(\ref{cond1Gzero}), and
\begin{equation}
r_2^{(n)}=-\frac{2\,\left[\arcsin\left(\frac{\mu}{\sqrt{4\,\omega^{2}+\mu^{2}}}\right)
                  -\pi\, n\right]}{\mu}\,,\; n=1,2,\ldots\;,
\label{eq:G trig zero r2n}
\end{equation}
\begin{equation}
r_{3}^{(n)}=\frac{2\,\left[\arcsin\left(\frac{\mu}{\sqrt{4\,\omega^{2}+\mu^{2}}}\right)
                  +\pi\, n\right]}{\mu},\; n=0,1,2,\ldots
\label{eq:G trig zero r3n}
\end{equation}
for Eqs.~(\ref{cond2Gzero}).

{}From equations~(\ref{eq:G trig zero r1n}), (\ref{eq:G trig zero r2n})
and~(\ref{eq:G trig zero r3n}) one has that although the values of the maxima,
minima and zeros of $G(r)$ change for different values of the parameters $\mu$
and $\omega$,  the general behavior of $G(r)$ (shape of the curve, number of maxima,
minima and zeros) does not depend on the specific values of $\mu$ and $\omega$.
Thus, for example, for $\mu=2$ and $\omega=1/2$ one has that
$G(r)=-\frac{1}{4}(\,5\,\sin^2r-4)\,\sin^2r$, whose graph is shown
in the Figure~\ref{fig1}. Different values of the parameters
$\mu$ and $\omega$ would give rise to a curve with the similar
global pattern but with different values for the minima, maxima and zeros.

Finally, from  the above results one obtains the sequence of alternating causal
[$\,G(r) > 0\,$] and noncausal [$\,G(r) < 0\,$] regions. Indeed,  $G(r)>0$
for
\begin{equation}
R_{1}=\left\{ r \mid  (r_1^{(0)}=0) \leq r \leq \: r_{3}^{(0)} =1.11\right\} \,,
\label{eq:trig R1 causal}
\end{equation}
\begin{equation}
R_{n}=\left\{r \mid r_{2}^{(n-1)} \leq r \leq \: r_{3}^{(n-1)}\right\},\;n=2,3,\ldots \,,
\label{eq:trig Rn causal}
\end{equation}
and $G(r) < 0 $ otherwise.

Finally, it should be noticed that the analysis carried out in this appendix fulfill
a minor gap left in Ref.~\cite{Reb_Tiomno} in the study of the trigonometric class
of G\"odel-type geometries.


\end{document}